# A Multiband T-Shaped Antenna Array for 6G Mobile Communication

Sunday Achimugu[1*], Abraham Usman Usman[2], Suleiman Zubair[3], Michael David[4], Abdulkadir Olayinka Abdulbaki[5], Hassan Musa Abdullahi[6]

[1,2,3,4,5,6]Department of Telecommunications Engineering
Federal University of Technology, Minna
*sundayachimugu@ieee.org

**ABSTRACT**

The paradigm shift in the use cases of wireless communication necessitates the need to move toward higher data rates, large bandwidths, and intelligent reconfiguration in 6G. This paper presents a novel double T-shaped antenna array that operates between 4GHz to 16 GHz for 6G mobile communication. The antenna consists of a rectangular microstrip with a fractal T-shaped slot, cut at the rear of the microstrip to provide an air gap for an improved radiation pattern. The antenna is separated at 9 degrees to create a $3 \times 3$ array which is fed through a $0.8 \times 2.77$ mm$^2$ microstrip line. The antenna was designed and simulated using the CST suite based on Finite Integration Technique. The antenna exhibits a peak gain of 7.99 dBi at 15.35 GHz, with a maximum reflection coefficient of -31.01 dBi at 8.43 GHz. It also operates in four bands, 6.62-7.54 GHz, 8.27-8.78 GHz, 10.98-11.78 GHz, and 13.65-15.4 GHz respectively, realizing a 92% radiation efficiency between 8.27-8.78 GHz frequency band. The antenna array was fabricated on Rogers 5880 substrate with 2.2 $\varepsilon$, and 0.766 mm thickness, displaying an $S_{11} \leq$ -10 dB over the entire operating frequency. This validates the simulated results. The antenna is compact ($30.9 \times 26.63 \times 0.1$ mm3) and suitable for 6G mobile broadband applications.

## 1. INTRODUCTION

THE emergence of wireless communication has paved the way for the seamless transmission of information. The huge demand for data rates has resulted in the transition and upgrade of mobile communication networks every ten years. Already in use, Fifth-generation (5G) wireless communication networks enable machines to communicate with each other, and deliver high data rates (Andersson et al., 2023). Nevertheless, 5G specifications have not comprehended the needs of the emerging technologies and use cases. As we move toward complex use cases including, connected intelligent machines, teleporting, ubiquitous connectivity, tactile internet, smart transport, ultra-high-definition videos, and Tbps data rates, 6G has been envisioned (Wang et al., 2023). This has led to a continued need for increased data capacity which in turn requires more bandwidth and spectrum. The next generation of mobile communication, 6G is rapidly receiving attention within the communication industry and academia considering its first commercial deployment which is planned for 2030. To advance the development of 6G technology, the ITU-R at WRC 2023 harmonized the frequency band for 6G mobile communication and advanced use case scenarios. For cellular mobile communication, 4400-4800MHz, 7125-7250MHz, 7750-8400MHz, and 14.80-15.35GHz have been allocated to region 1. Region 2 is to utilize 7125-8400MHz and 14.80-15.35GHz bands. While Region 3 is to use 4400-4800MHz, 7125-8400MHz, and 14.80-15.35GHz, all regions have been licensed to use the 6425MHz – 7125MHz frequency band. In addition, millimeter-wave (mm-Wave) and sub-THz bands within 102-275GHz (102-109 GHz, 151.5-164 GHz, 167-174 GHz, 209-226 GHz 252-275 GHz) have been earmarked for 6G advanced communication scenarios (Castro, 2023).

The design of antennas that can effectively operate across these frequency bands is pivotal to deploying 6G. The antennas used in cellular and Wi-Fi systems have gradually moved from simple patch, dipole, and monopole antennas in the early generation of wireless communication to advanced beamforming array antennas used in the multiple-input multiple-output (MIMO) systems in the fourth (4G) and the latest fifth (5G) generations of wireless technologies (Achimugu et al., 2023), (Olwal et al. 2021). In 5G mobile communication systems, a lower mm-wave band (<100GHz) is sufficient to deliver 10Gbps data rate (Vyas, 2019). For such a band, multiple-input-multiple-output (MIMO) antenna configurations are been widely implemented. The authors (Griffiths et al., 2023) presented a mm-wave antenna design using a ring-slot resonator to achieve a wide bandwidth of 25.0-29.7 GHz.

Although MIMO antennas deliver good spatial radiation, the coupling between the ports is a major concern as it degrades the performance of the antenna. As a result, several attempts have been undertaken to increase the isolation between radiators. One of the ways to achieve good isolation in MIMO antennas is to use a metamaterial-based MIMO design (You et al. 2021). MIMO technology has been widely utilized in mobile cellular antennas to allow operation at different bands. Four elements coplanar-fed array antenna operating across 2-5.2 GHz was proposed by (Hassan et al., 2023) for 5G mobile communication. The antenna realized a peak gain of 6.2dBi at 3.8 GHz. Creating an air gap between the design elements of an antenna is a means of achieving magnetic and electric coupling, providing isolation and improving radiation efficiency. (Naqvi & Hussain, 2022). (Hassan et al., 2023) provided an air gap of 1mm and 6mm between a front triangular-shaped array and the fractal back in their design. This provided an improved isolation and reflection coefficient of the antenna.

In a related work, (Tuib et al., 2023) presented a 2 by-2 circular patch MIMO antenna array operating between 26.5-29.5GHz. The individual antenna in the array has a measured impedance bandwidth of 1.6 GHz from 27.25 to 28.85 GHz for $S11 \leq -10$ dB, while the MIMO array delivered a gain of 7.2 dBi at 28 GHz with inter radiator isolation greater than 26 dBi. Introducing frequency selective surfaces (FSS) is another way to improve the performance of antennas. The gain of the MIMO array was increased when (Tuib et al., 2023) introduced an FSS consisting of a $7 \times 7$ array of rectangular C-shaped resonator unit cells. With the FSS, the gain of the MIMO array increased to 8.6 dB at 28 GHz.

(Baz et al., 2023) suggested a circularly slotted $4 \times 4$ MIMO antenna. The antenna resonates at three different bands, 2.02 GHz, 5.87 GHz, and 11.19 GHz, realizing a maximum gain of 10.94 dB at 3.94 GHz, and was fabricated on a low-profile FR4 substrate.

This paper presents a novel double T-shaped multiband-array antenna for mobile communication in 6G technology. The antenna operates across a wide bandwidth of 12GHz, suitable for application in 6G mobile broadband devices. The rest of the work is arranged as follows. In section 2, the structural design of the antenna is presented. The results are discussed in Section 3. In section 4, a conclusion is drawn while section 5 is acknowledgment.

2. **ANTENNA GEOMETRY AND DESIGN**

A double T-shaped patch antenna fabricated on Rogers 5880 (RT) substrate with permittivity (ε) of 2.2, thickness h, and 0.00009 loss tangent has been developed following four stages. RT dielectric was selected because of its low $ε_r$, relatively low loss tangent, and ability to withstand tear. The proposed antenna structure design stages are shown in Figure 1. The antenna operates between 4 to 16 GHz, covering the four bands allocated for 6G mobile communication. Stage 1 represents a rectangular patch structure made of copper, designed following (1) to (7) respectively. The novelty of the design is the double fractal T-shaped strategically cut in the second stage as shown in Figure 1(b). The T-shaped provides an air gap between the substrate and the radiating copper, resulting in an improvement of the bandwidth and radiation efficiency of the antenna. In stage 3, the T-shaped fractal structure is made into a 3 by 3 array separated at an angle of 9 degrees from each other. This allows the antenna to operate over four frequency bands.

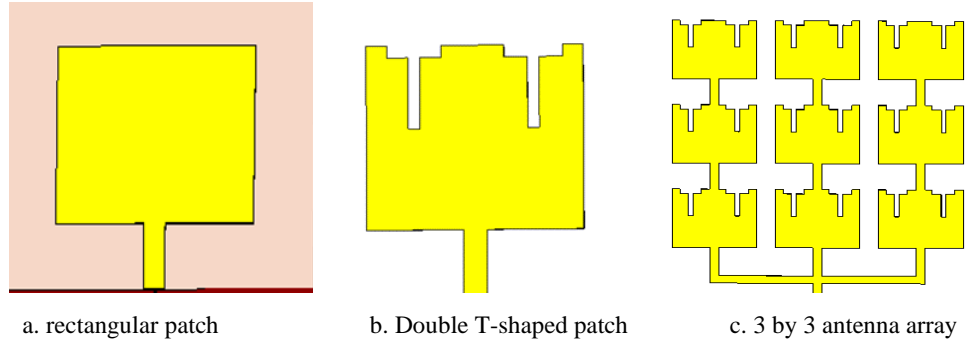

a. rectangular patch   b. Double T-shaped patch   c. 3 by 3 antenna array

**Figure 1:** Evolution of the Proposed Radiating Antenna Structure

The Length of the rectangular microstrip antenna is represented by $L$ while $\lambda_0$ is the wavelength of the antenna's resonant frequency, $f_r$. The extension of the patch is represented by $\Delta l$ and derived from equation 4. $\varepsilon_e$ is the effective permittivity of the material and $c$ is the speed of light. The length and width of the substrate for optimal radiation are obtained from (1) and (2) respectively, with $f_r = 6$ GHz. The final proposed antenna structure is shown in Figure 5. The optimized physical parameters of the antenna are listed in Table 1.

$$L = \frac{\lambda_0}{2\sqrt{\varepsilon_e}} - \Delta l \quad (1)$$

The width of the patch is obtained from (2)

$$W = \frac{c}{2f_r}\left(\frac{1}{\sqrt{\left(\frac{\varepsilon_r+1}{2}\right)}}\right) \quad (2)$$

The effective permittivity $\varepsilon_e$ is obtained from (3),

$$\varepsilon_e = \frac{e_r+1}{2} + \frac{e_r-1}{2}\left\{\frac{1}{\sqrt{1+\frac{12h}{W}}}\right\} \quad (3)$$

$$\Delta l = 0.412h \left(\frac{\varepsilon_e+0.3}{\varepsilon_e-0.258}\right)\left(\frac{(W/h)+0.264}{(W/h)+0.8}\right) \quad (4)$$

$$L_s = L + 6h \quad (5)$$

$$W_s = W + 6h \quad (6)$$

$$h = \frac{0.0606\lambda_0}{\sqrt{e_r}} \quad (7)$$

**Table 1:** Antenna Structural Design Parameters

| SYMBOL | PARAMETER | UNIT (mm) |
|---|---|---|
| L | Length of the patch | 6.23 |
| W | Width of the patch | 7.41 |
| $f_g$ | Feed line gap | 0.80 |
| $f_l$ | Feed line length | 2.77 |
| $t_a$ | T-shaped arm | 0.72 |
| $t_g$ | T-shaped width | 0.4 |
| $t_l$ | T-shaped length | 2.4 |
| $t_b$ | T-shaped breadth | 0.48 |
| d | T separation length | 2.16 |
| g | Array gap | 2.2 |

## 3. RESULT AND DISCUSSION

### 3.1. Rectangular Patch

Figure 2 (a) shows the reflection coefficient of the rectangular patch in Figure 1 (a). A maximum performance was realized at the 8.25 GHz band. The antenna has a peak gain of 6.38 dB at the 15.35 GHz band. The large bandwidth of $S_{11} \geq -10$dB exhibited by the antenna indicates high distortion of the antenna and hence poor performance.

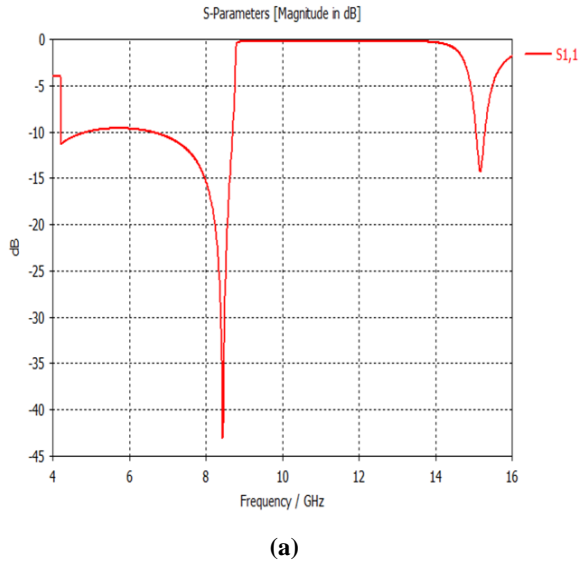
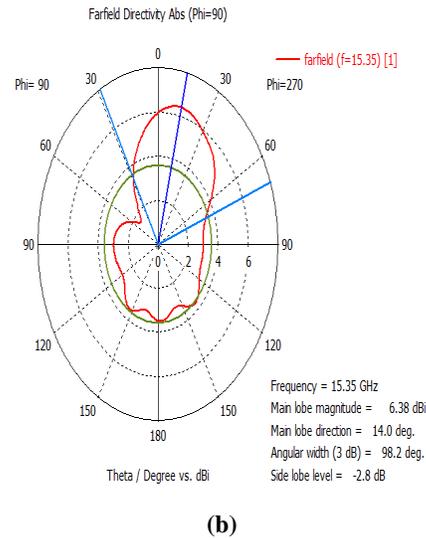

(a)  (b)

**Figure 2:** (a) $S_{11}$ of Single Patch Antenna (b) Peak Gain of Single Patch Antenna

### 3.2. Double T-shaped Antenna

To improve the performance of the results in Figure 2, a double 1.84 × 2.4 × 2.48 mm3 T-slot cut is strategically made at the top of the patch antenna as displayed in Figure 3(a). The double T-shaped slots enhanced the performance of the antenna as can be seen in Figure 4. Also, a reduced side lobe of -8.8dBi, and a maximum gain of 6.63 dBi was obtained at the 14.8 GHz band. However, the large bandwidth of S11 ≥ -10dB exhibited by the antenna is not desirable.

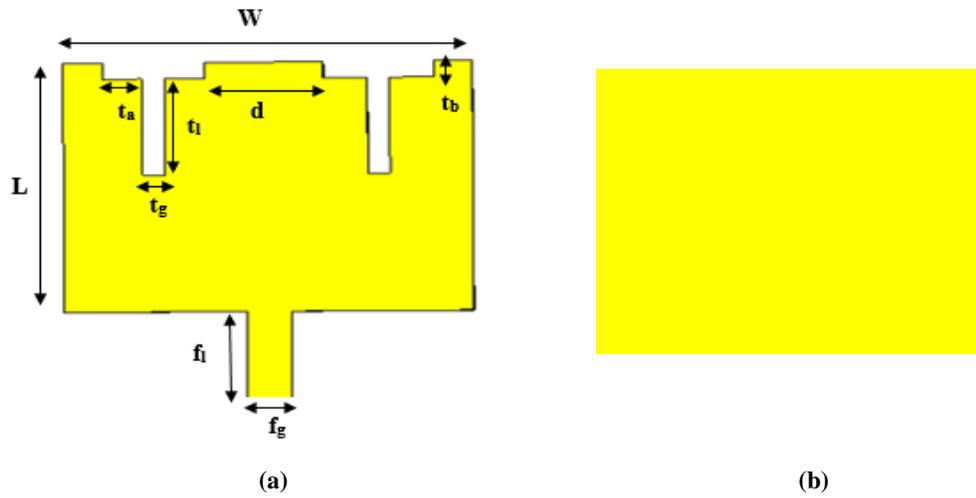

**Figure 3:** (a) Front View of Novel Double T-shaped Antenna (b) Back-view of T-shaped Antenna

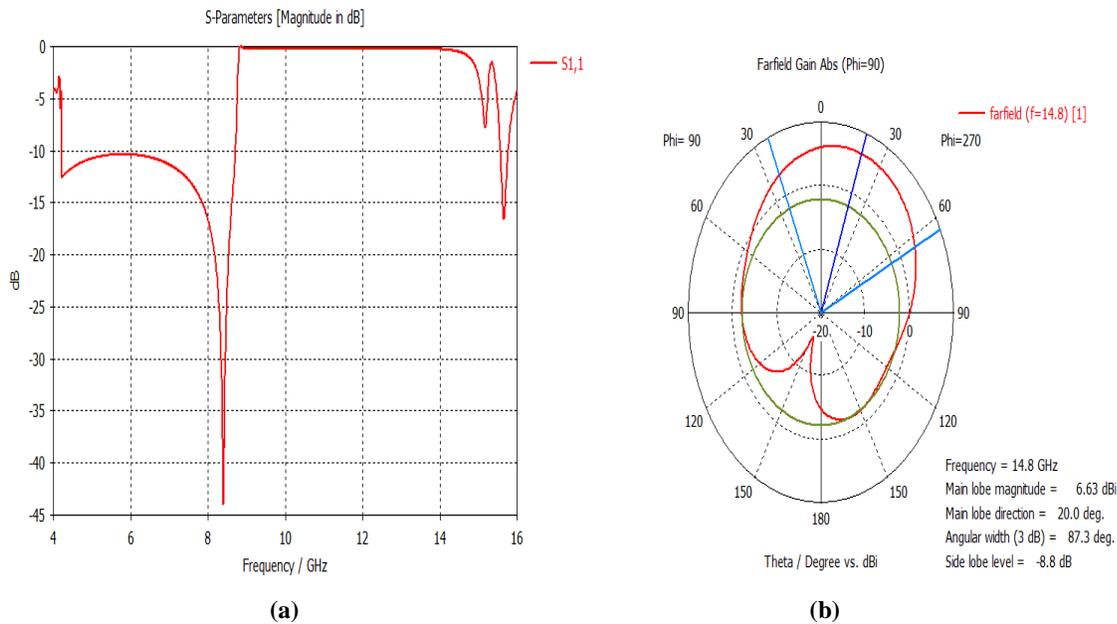

**Figure 4:** (a) $S_{11}$ of Double T-shaped Antenna (b) Peak Gain of Double T-shaped Antenna

### 3.3. 3 × 3 Antenna Array

The double T-shaped fractal slot rectangular patch antenna was formed into a 3 × 3 antenna array. The individual antennas in the array were spatially arranged orthogonally at 9 degrees to each other with a 1.59 mm horizontal gap, as shown in Figure 5(a) to improve the isolation between the individual antennas. A $0.8 \times 0.78$ mm$^2$ coplanar waveguide microstrip feed line of 50 Ω input impedance was used for the input port. The overall dimension of the array (l × w × t) as shown in Figure 5 is $30.9 \times 26.63 \times 0.1$ mm$^3$. The simulation results in Figure 6 show the antenna's improved performance. The antenna exhibits a peak gain of 7.99 dB at 15.35 GHz, with a maximum reflection coefficient of -31.01 dB at 8.43 GHz.

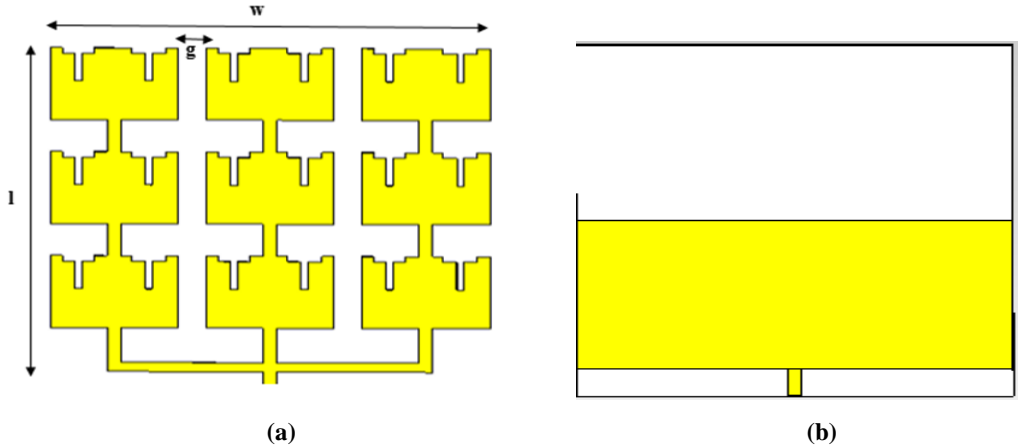

**Figure 5:** (a) Front view 3 × 3 antenna array (b) Back view 3 × 3 antenna array

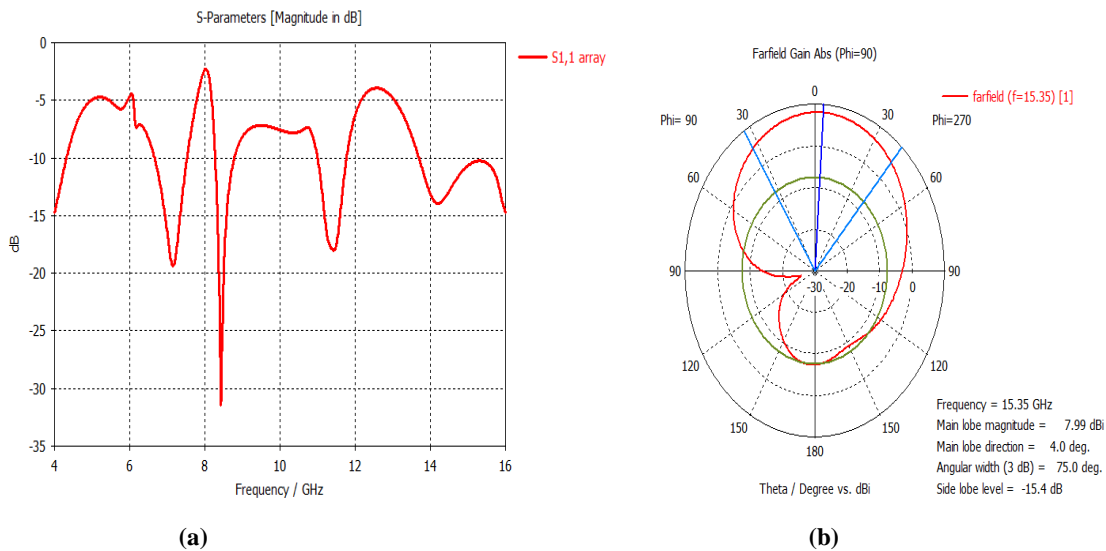

**Figure 6:** (a) $S_{11}$ of 3 × 3 Antenna Array (b) Peak Gain of 3 × 3 Antenna Array

6G mobile communication antenna will operate in multiple bands, with high directivity (Naqvi & Hussain, 2022). From Figure 6(a), it can be seen that the antenna operates in four bands, where $S_{11} \leq -10$ dB. These correspond to 6.62-7.54 GHz, 8.27-8.78 GHz, 10.98-11.78 GHz, and 13.65-15.31 GHz bands respectively. This makes the antenna suitable for 6G mobile communication. The antenna array was fabricated on Rogers 5880 substrate with 2.2 $\varepsilon$, and 0.766 mm thickness as shown in Figure 7. The antenna displayed an $S_{11} \leq -10$ dB over the entire operating frequency. This validates the simulated results. A comparison of the antenna proposed in this paper to other similar work is presented in Table 2.

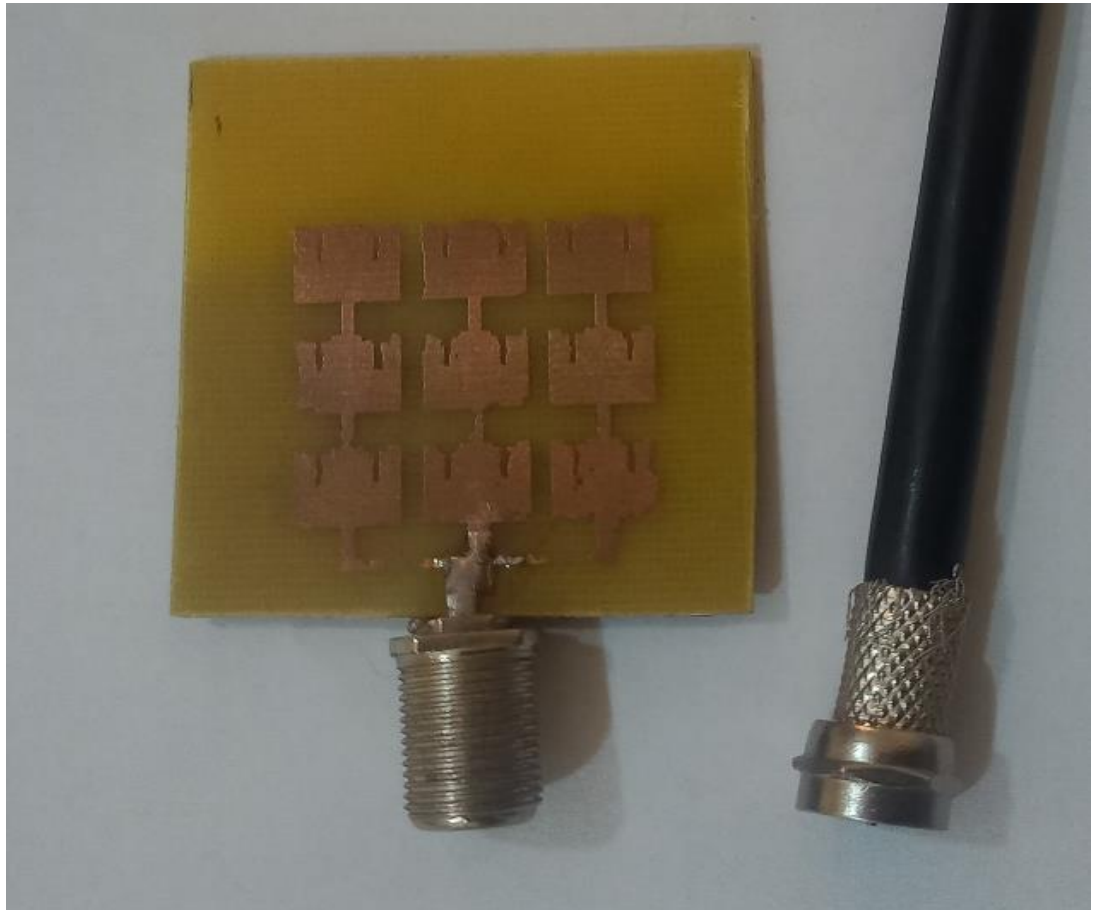

**Figure 7:** Fabricated 3 × 3 Antenna Array

**Table 2:** Comparison of proposed antenna to other related work

| PAPER | FREQUENCY (GHz) | ANTENNA TYPE | GAIN (dB) | BANDWIDTH (GHz) | REFLECTION COEFFICIENT (dB) |
|---|---|---|---|---|---|
| (Wong et al., 2023) | 7.025-8.4 | 2 × 2 Array | 7.35 | 1.375 | --22 |
| (Abdulhussein et al., 2023) | 2.7-4.8 | Microstrip patch | - | 2.1 | -32.33 |
| (Ntawangaheza et al., 2020) | 4.48-6.8 | 4 × 4 array | 10.14 | 2.32 | -19.8 |
| (Sayem et al., 2023) | 2.4-4.29 | Microstrip patch | 7.24 | 1.89 | -47.21 |
| **This paper** | **4-16** | **3 ×3 Array** | **7.99** | **12** | **-31.01** |

## 4. CONCLUSION

This paper presents a novel double T-shaped multiband antenna for 6G mobile communication. The antenna operates between 4GHz to 16GHz. A peak gain of 7.99 dB is obtained at 15.35GHz. Simulation results show that the antenna operates in four bands including, 6.62-7.54 GHz, 8.27-8.78 GHz, 10.98-11.78 GHz, and 13.65-15.4 GHz respectively. The antenna array was fabricated on Rogers 5880 substrate with 2.2 $\varepsilon$, and 0.766 mm, displaying an $S_{11} \leq$ -10 dB over the entire operating frequency. This validates the simulated results. The antenna is compact and suitable for 6G mobile applications requiring operation in different bands.

## ACKNOWLEDGEMENT

The authors acknowledge the Nigerian Communication Commission that sponsored this work under the Professorial Endowment Fund, awarded to the Federal University of Technology Minna.